\documentclass[twoside,twocolumn,9pt]{article}

\usepackage[utf8]{inputenc}

\usepackage{extsizes}
\usepackage[super,sort&compress,comma]{natbib}

\usepackage{chemformula}
\let\ce\ch

\usepackage[left=1.5cm, right=1.5cm, top=1.785cm, bottom=2.0cm]{geometry}
\usepackage{balance}
\usepackage{mathptmx}
\usepackage{sectsty}
\usepackage{graphicx} 
\usepackage{lastpage}
\usepackage[format=plain,justification=justified,singlelinecheck=false,font={stretch=1.125,small,sf},labelfont=bf,labelsep=space]{caption}
\usepackage{float}
\usepackage{fancyhdr}
\usepackage{fnpos}
\usepackage[english]{babel}
\addto{\captionsenglish}{%
  
}
\usepackage{array}
\usepackage{droidsans}
\usepackage{charter}
\usepackage[T1]{fontenc}
\usepackage{xcolor}
\usepackage{setspace}
\usepackage[compact]{titlesec}
\usepackage{hyperref}
\usepackage{color,soul}

\usepackage{epstopdf}

\definecolor{cream}{RGB}{222,217,201}

\usepackage{gensymb}

\usepackage{booktabs}
\usepackage{siunitx}
\DeclareSIUnit{\mbar}{mbar}
\DeclareSIUnit{\ppm}{ppm}
\DeclareSIUnit{\molec}{molec}
\DeclareSIUnit{\mPa}{mPa}

\begin{document}

\newcommand{\Agilio}[1]{\textcolor{purple}{#1}}

\pagestyle{fancy}
\thispagestyle{plain}
\fancypagestyle{plain}{
\renewcommand{\headrulewidth}{0pt}
}

\makeFNbottom
\makeatletter
\renewcommand\LARGE{\@setfontsize\LARGE{15pt}{17}}
\renewcommand\Large{\@setfontsize\Large{12pt}{14}}
\renewcommand\large{\@setfontsize\large{10pt}{12}}
\renewcommand\footnotesize{\@setfontsize\footnotesize{7pt}{10}}
\makeatother

\renewcommand{\thefootnote}{\fnsymbol{footnote}}
\renewcommand\footnoterule{\vspace*{1pt}%
\color{cream}\hrule width 3.5in height 0.4pt \color{black}\vspace*{5pt}} 
\setcounter{secnumdepth}{5}

\makeatletter 
\renewcommand\@biblabel[1]{#1}            
\renewcommand\@makefntext[1]%
{\noindent\makebox[0pt][r]{\@thefnmark\,}#1}
\makeatother 
\renewcommand{\figurename}{\small{Fig.}~}
\sectionfont{\sffamily\Large}
\subsectionfont{\normalsize}
\subsubsectionfont{\bf}
\setstretch{1.125} 
\setlength{\skip\footins}{0.8cm}
\setlength{\footnotesep}{0.25cm}
\setlength{\jot}{10pt}
\titlespacing*{\section}{0pt}{4pt}{4pt}
\titlespacing*{\subsection}{0pt}{15pt}{1pt}

\fancyfoot{}
\fancyfoot[LO,RE]{\vspace{-7.1pt}\includegraphics[height=9pt]{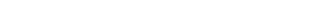}}
\fancyfoot[CO]{\vspace{-7.1pt}\hspace{13.2cm}\includegraphics{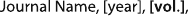}}
\fancyfoot[CE]{\vspace{-7.2pt}\hspace{-14.2cm}\includegraphics{head_foot/RF}}
\fancyfoot[RO]{\footnotesize{\sffamily{1--\pageref{LastPage} ~\textbar  \hspace{2pt}\thepage}}}
\fancyfoot[LE]{\footnotesize{\sffamily{\thepage~\textbar\hspace{3.45cm} 1--\pageref{LastPage}}}}
\fancyhead{}
\renewcommand{\headrulewidth}{0pt} 
\renewcommand{\footrulewidth}{0pt}
\setlength{\arrayrulewidth}{1pt}
\setlength{\columnsep}{6.5mm}
\setlength\bibsep{1pt}

\makeatletter 
\newlength{\figrulesep} 
\setlength{\figrulesep}{0.5\textfloatsep} 

\newcommand{\topfigrule}{\vspace*{-1pt}%
\noindent{\color{cream}\rule[-\figrulesep]{\columnwidth}{1.5pt}} }

\newcommand{\botfigrule}{\vspace*{-2pt}%
\noindent{\color{cream}\rule[\figrulesep]{\columnwidth}{1.5pt}} }

\newcommand{\dblfigrule}{\vspace*{-1pt}%
\noindent{\color{cream}\rule[-\figrulesep]{\textwidth}{1.5pt}} }

\makeatother

\twocolumn[
  \begin{@twocolumnfalse}
{\includegraphics[height=30pt]{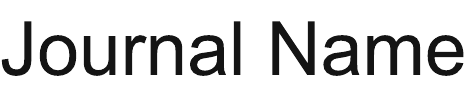}\hfill\raisebox{0pt}[0pt][0pt]{\includegraphics[height=55pt]{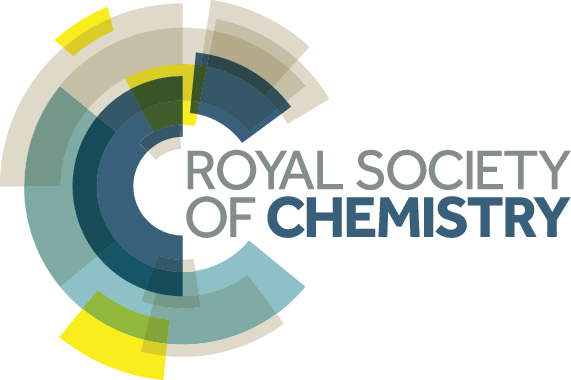}}\\[1ex]
\includegraphics[width=18.5cm]{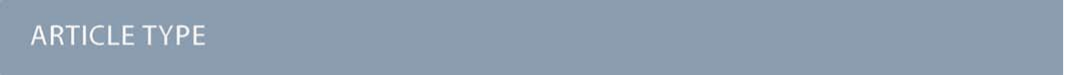}}\par
\vspace{1em}
\sffamily
\begin{tabular}{m{4.5cm} p{13.5cm} }

\includegraphics{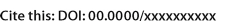} & \noindent\LARGE{\textbf{Effect of Ion Structure on the Physicochemical Properties and Gas Absorption of Surface Active Ionic Liquids$^\dag$}} \\
\vspace{0.3cm} & \vspace{0.3cm} \\

 & \noindent\large{Jocasta Avila,\textit{$^{a}$} Daniel Lozano-Martín,\textit{$^{b}$}, Mirella Simoes Santos,\textit{$^{a}$} Yunxiao Zhang,{$^{c}$}, Hua Li,{$^{c}$} Agilio Pádua,\textit{$^{a}$} Rob Atkin,{$^{c, \ast}$} and Margarida Costa Gomes\textit{$^{a,\ast}$}} \\

\includegraphics{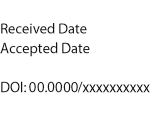} & \noindent\normalsize{

Surface active ionic liquids (SAILs) combine useful characteristics of both ionic liquids (ILs) and surfactants, hence are promising candidates for a wide range of applications. However, the effect of SAIL ionic structures on their physicochemical properties remains unclear, which limits their uptake. To address this knowledge gap, in this work we investigated the density, viscosity, surface tension, and corresponding critical micelle concentration in water, as well as gas absorption of SAILs with a variety of cation and anion structures. SAILs containing anions with linear alkyl chains have smaller molar volumes than those with branched alkyl chains, because linear alkyl chains are interdigitated to a greater extent, leading to more compact packing. This interdigitation also results in SAILs being about two orders of magnitude more viscous than comparable conventional ILs. SAILs at the liquid-air interface orient alkyl chains towards the air, leading to low surface tensions closer to n-alkanes than conventional ILs.  Critical temperatures of about \SI{900}{\K} could be estimated for all SAILs from their surface tensions. When dissolved in water, SAILs adsorb at the liquid-air interface and lower the surface tension, like conventional surfactants in water, after which micelles form. Molecular simulations show that the micelles are spherical and that lower critical micelle concentrations correspond to the formation of aggregates with a larger number of ion pairs. \ce{CO2} and \ce{N2} absorption capacities are examined and we conclude that ionic liquids with larger non-polar domains absorb larger quantities of both gases. 
} 

\end{tabular}

 \end{@twocolumnfalse} \vspace{0.6cm}

]

\renewcommand*\rmdefault{bch}\normalfont\upshape
\rmfamily
\section*{}
\vspace{-1cm}


\footnotetext{\textit{$^{a}$~Laboratoire de Chimie de l’ENS Lyon, CNRS and Université de Lyon, 46 allée d’Italie, 69364 Lyon, France. *margarida.costa-gomes@ens-lyon.fr}}
\footnotetext{\textit{$^{b}$~Grupo de Termodinámica y Calibración (TERMOCAL), Research Institute on Bioeconomy, Escuela de Ingenierías Industriales, Universidad de Valladolid, Paseo del Cauce, 59, 47011 Valladolid, Spain}}
\footnotetext{\textit{$^{c}$~School of Molecular Sciences, The University of Western Australia, Perth, Western Australia. *rob.atkin@uwa.edu.au}}

\footnotetext{\dag~Electronic Supplementary Information (ESI) available: [Experimental detail and numerical data in density, viscosity, surface tension and gas absorption measurements]. See DOI: 00.0000/00000000.}



\section{Introduction}

Ionic liquids (ILs) are pure salts with low melting temperatures\cite{Wang2020, Hayes2015} that have remarkable physicochemical properties, including high thermal and chemical stabilities, low flammability and volatility, wide electrochemical windows, and good solubility towards both polar and non-polar solutes.\cite{Silvester2021, Zhou2017b} Physicochemical properties can be tuned by mixing-and-matching cations and anions with different structures, such as different alkyl chain lengths. For instance, IL viscosity generally increases with alkyl chain lengths,\cite{Khalil2020} whereas density and surface tension decrease.\cite{Filippov2014, Tariq2012} The tunability of IL physicochemical properties is at least partially a consequence of the heterogeneous nanostructure of many ILs. Electrostatic interactions between charged centres cause them to cluster together into polar domains, from which alkyl chains are segregated solvophobically into apolar (uncharged) domains.\cite{Zhang2022} Tunable physicochemical properties make ILs promising candidates for a variety of applications including sensors,\cite{Silvester2019} catalysts,\cite{Choudhary2018} electrolytes,\cite{Doblinger2020} lubricants,\cite{Cooper2019} and, notably for this work, gas adsorbents. ILs have drawn great attention as favourable absorbents for anthropogenic emissions, such as \ce{CO2}, \ce{SOx}, and \ce{NOx}.\cite{Avila2021d, Kunov-Kruse2016, Zhang2009} IL gas solubility can be further enhanced by suspending metal-organic frameworks (MOFs) in ILs to form porous ILs.\cite{Avila2021, Avila2021d, CostaGomes2018c}

Surface active ionic liquids (SAILs) are a subclass of ILs that usually contain an amphiphilic ion composed of a polar charged centre and long apolar alkyl chain tails.\cite{Brown2012a, Brown2011} SAILs reported to date have surfactant based anions, such as 1,4-bis(2-ethylhexoxy)-1,4-dioxobutane-2-sulfonate (\ce{[AOT]-}) and dodecyl sulfate (\ce{[DS]-}). SAILs combine some useful characteristics of both ILs and surfactants, and self-assemble into very well defined bulk nanostructures reminiscent of aqueous surfactant sponge phases.\cite{Hayes2015, Zhang2009, Sharma2021} 
The lengths of alkyl chains affect the dimensions of the bulk nanostructure, resulting in an adaptable capacity to dissolve and extract with, for example, biomaterials\cite{Ke2018,Jin2016,Mao2019}. Our recent work reveals that the bulk nanostructures of SAILs can change from anion bilayer structures to cation-anion interdigitated structures as the ion structures change from short alkyl chain cations and linear alkyl chain anions to long alkyl chain cations and branched alkyl chain anions \cite{Zhang2022,Zhang2022a}.
SAILs are also better lubricants than conventional ILs because the long surfactant alkyl chains form a robust boundary layer that reduces energy dissipation and thus friction.\cite{Zhang2022} 

As SAILs contain amphiphilic ions, when dissolved in water or in organic solvents, they form self-assembled systems similar to conventional surfactants, such as micelles and vesicles,\cite{Banerjee2022, Lepori2020} which have shown potential for applications in biomedicine\cite{Lepori2019} or drug delivery.\cite{Pattni2015a} Research on SAILs has focused so far on systems where a surfactant anion is combined with an imidazolium cation. Only recently, the bulk and interfacial structure of phosphonium and ammonium based SAILs have been investigated \cite{Zhang2022,Zhang2022a}. The effect of the structure of the cation, the nature of the anion head group and of the alkyl chain length, on physiochemical properties of the ionic liquids is unclear, which restricts the rational design of SAILs. Similarly, gas absorption by SAILs has received little attention, although they are promising gas absorbents due to their excellent stability and tunability. To address these knowledge gaps, in this study a series of SAILs with imidazolium, phosphonium, and ammonium cations as well as \ce{[AOT]-} and \ce{[DS]-} anions are studied. Physicochemical properties, including density, viscosity, surface tension, and critical micelle concentrations (cmc) are characterised. The effect of the SAIL ionic structure on absorption capacities for \ce{N2} and \ce{CO2} is examined experimentally. The outcomes of this work provide a deep understanding of how SAIL ionic structures tune their physicochemical properties, which will enable SAIL ionic structures to be rationally designed for a given application.

\section{Experimental}

\subsection{Materials}

The ionic liquids \textbf{(a)} 1-butyl-3-methylimidazolium  dioctyl sulfosuccinate - \ce{[C4mim][AOT]} --- \textbf{(b)} tetrabutylammonium dioctyl sulfosuccinate - \ce{[N_{4,4,4,4}][AOT]} --- \textbf{(c)} tetrabutylphosphonium dioctyl sulfosuccinate - \ce{[P_{4,4,4,4}][AOT]} --- \textbf{(d)} trihexyltetradecylphosphonium dioctyl sulfosuccinate - \ce{[P_{6,6,6,14}][AOT]} --- \textbf{(e)} tetrabutylammonium dodecyl sulfate - \ce{[N_{4,4,4,4}][DS]} --- \textbf{(f)} tetrabutylphosphonium dodecyl sulfate - \ce{[P_{4,4,4,4}][DS]} --- \textbf{(g)} trihexyltetradecylphosphonium dodecyl sulfate - \ce{[P_{6,6,6,14}][DS]} - were synthesized and characterized as described in the Supplementary Information. Fig.\ref{fig:Struc_SAILs} shows the chemical structures of the seven surfactant ionic liquids studied herein. Before utilization, the salts, that are liquid at room temperature, were degassed for at least 72 hours under a primary vacuum and were kept dried and degassed inside a desiccator. The water content of the ionic liquids was determined by the Karl Fischer coulometric titration method (Mettler Toledo C20S with reagent Hydranal Coulomat E), the values measured being listed in Table S1.

Water HPLC was purchased from Sigma Aldrich and was used without further purification.

Carbon dioxide, \ce{CO2} 4.5, and nitrogen, \ce{N2} 4.5, were purchased from Air Liquide with a mole fraction purity of \SI{99.995}{\percent} and were used as received.

The ILs/water mixtures listed in Table S9 were prepared gravimetrically using a New Classic MS Mettler Toledo balance with an accuracy of \SI{\pm0.01}{\mg}. The estimated uncertainty on the ionic liquid mole fraction concentration is \SI{\pm0.0005}{\percent}.

\begin{figure}
\begin{center}
\includegraphics[width=10cm]{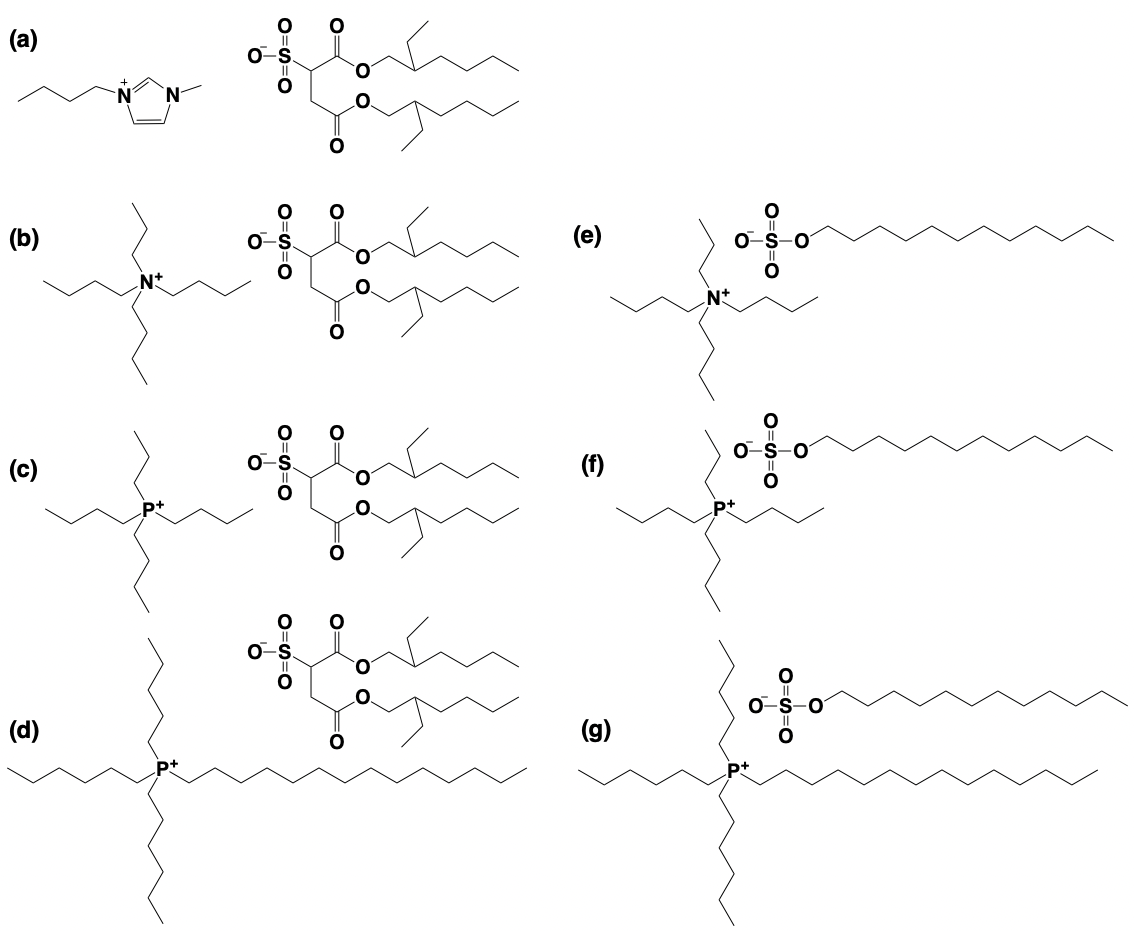}
\end{center}
\caption{Surfactant ionic liquids studied in this work: (a) \ce{[C4mim][AOT]}, (b) \ce{[N_{4,4,4,4}][AOT]}, (c) \ce{[P_{4,4,4,4}][AOT]}, (d) \ce{[P_{6,6,6,14}][AOT]}, (e) \ce{[N_{4,4,4,4}][DS]}, (f) \ce{[P_{4,4,4,4}][DS]}, (g) \ce{[P_{6,6,6,14}][DS]}.}
\label{fig:Struc_SAILs}
\end{figure}

\subsection{Density and viscosity measurements}

The densities of the samples were measured in an Anton Paar DMA 5000M densimeter, in the temperature range \SIrange{293}{353}{\K} at atmospheric pressure. The densimeter was calibrated before the measurements with two substances of accurately known densities: air and ultra-pure water provided as reference substance by Anton Paar. Temperature is controlled to within \SI{\pm0.001}{\celsius}.

Rheological measurements were conducted using a TA Instruments DHR-3 rheometer with an attached Peltier plate. A stainless-steel geometry with a cone of \SI{2}{\degree} and a diameter of \SI{20}{\mm} was used. The temperature was increased from \SI{283}{\K} to \SI{333}{\K} at a fixed low shear rate of \SI{50}{\per\second}. All runs were performed in triplicate with the replacement of IL samples between each run. Errors were within $10\%$.

\subsection{Surface tension measurements}

Air-ionic liquid surface tension, $\gamma$, was measured by the pendant drop method on a Krüss DSA30 drop shape analyzer at temperatures between \SIrange{293}{353}{\K} and at atmospheric pressure. The technique is based on recording the profile of a pendant drop from the tip of an injection needle by means of a CCD camera and a low-magnification lens system. 




The software numerically fits the Bashforth-Adams equation \cite{Bashforth1883} to the drop profile for each frame recorded by the camera, provided that an optical calibration of the instrument was carried out. The experimental procedure consists of performing small liquid injections, while simultaneously recording the pendant drop during its growth, discarding those values that show a dependency with the volume of the drop, for several repetitions at each temperature. 
Temperature of the needle chamber is controlled by a Grant GR150 thermal bath and measured by the temperature probe of the DSA30 instrument with an accuracy of \SI{0.1}{\celsius}. The experimental expanded ($k = 2$) uncertainty of the surface tension data is estimated to be not worse than $2 \%$, which is about five times the repeatability of the measurements given by the standard deviation between repetitions. 

\subsection{Gas absorption measurements}

Gas solubilities were measured by a gravimetric method using an Intelligent Gravimetric Analyzer (IGA001) from Hiden Analytical in the \SIrange{0.5}{5}{\bar} pressure range at \SI{303}{\K}. The working principle and the data treatment are extensively discussed in our previous works.\cite{Avila2021, Avila2021c} Essentially, the mass of gas absorbed, $m_g$, at each pressure and temperature was obtained from the raw weight data, $m_\mathrm{reading}$, using equation~(\ref{eq:mreading}).
\begin{equation}
    \label{eq:mreading}
    \begin{split}
    m_\mathrm{reading} &= m_0 +  m_s + m_g + m_g^\mathrm{EP} \\
    &- \sum_i \frac{m_i}{\rho_i} \rho_g(T_i,p) + \sum_j \frac{m_j}{\rho_j} \rho_g(T_j,p) - \frac{m_s}{\rho_s(T_s)} \rho_g(T_s,p)
    \end{split}
\end{equation}
where $m_s$ is the mass of degassed sample, $m_g^\mathrm{EP}$ the effect due to adsorbed gas on the balance components (determined by performing a blank measurement), and the sums over the $i$ and $j$ components account for the respective buoyancy effects, on the sample and counterweight sides, respectively.

\subsection{Coarse-grained simulations}

Coarse-grained molecular dynamics simulations were performed to investigate the shape and size of micelles formed by the novel ionic liquid based-surfactants in aqueous solutions and, for that, the GROMACS \cite{Hess2008} software was used.  The Martini 2.0 force field \cite{Marrink2007} was used, where water was described by the P4 bead, and the ions as proposed elsewhere \cite{Schaeffer2019,Negro2014,Peroukidis2020}. Each simulation consisted of 0.1M aqueous solutions, with 60000 P4 beads and 432 ion pairs. An NPT ensemble was considered, where the temperature was kept at 298.15 K and the pressure at 1 bar. The Parrinello-Rahman barostat was used with a constant of 12.0, and the v-rescale thermostat with a constant of 1.0.  

The initial structure of the simulation boxes was obtained by randomly placing the ions and water beads within the box. After energy minimization, the equilibration was performed in three steps. The first one consisted of a \SI{2}{\ns} run, with a \SI{0.002}{\ps} timestep and the Coulombic interactions being treated with the Reaction-field formalism. Afterwards, the timestep is changed to \SI{0.005}{\ps}, and the simulation is ran for \SI{5}{\ns}. Finally, on the final equilibration stage, the timestep is \SI{0.001}{\ps}, the simulation is ran for \SI{10}{\ns}, and the Coulombic interactions are treated by the Particle-Mesh Ewald method. Finally, the production is performed with a timestep of \SI{0.003}{\ps}, which ran for \SI{300}{\ns}. We performed the analysis presented here on the last \SI{50}{\ns} of the simulations. We considered the cut-off approach for the van der Waals interactions, and the cut-off for these and the Coulombic interactions was \SI{0.11}{\nm}. 

\section{Results and Discussion}

\begin{figure}[hbt!]
    \centering
    \includegraphics[width=0.5\textwidth]{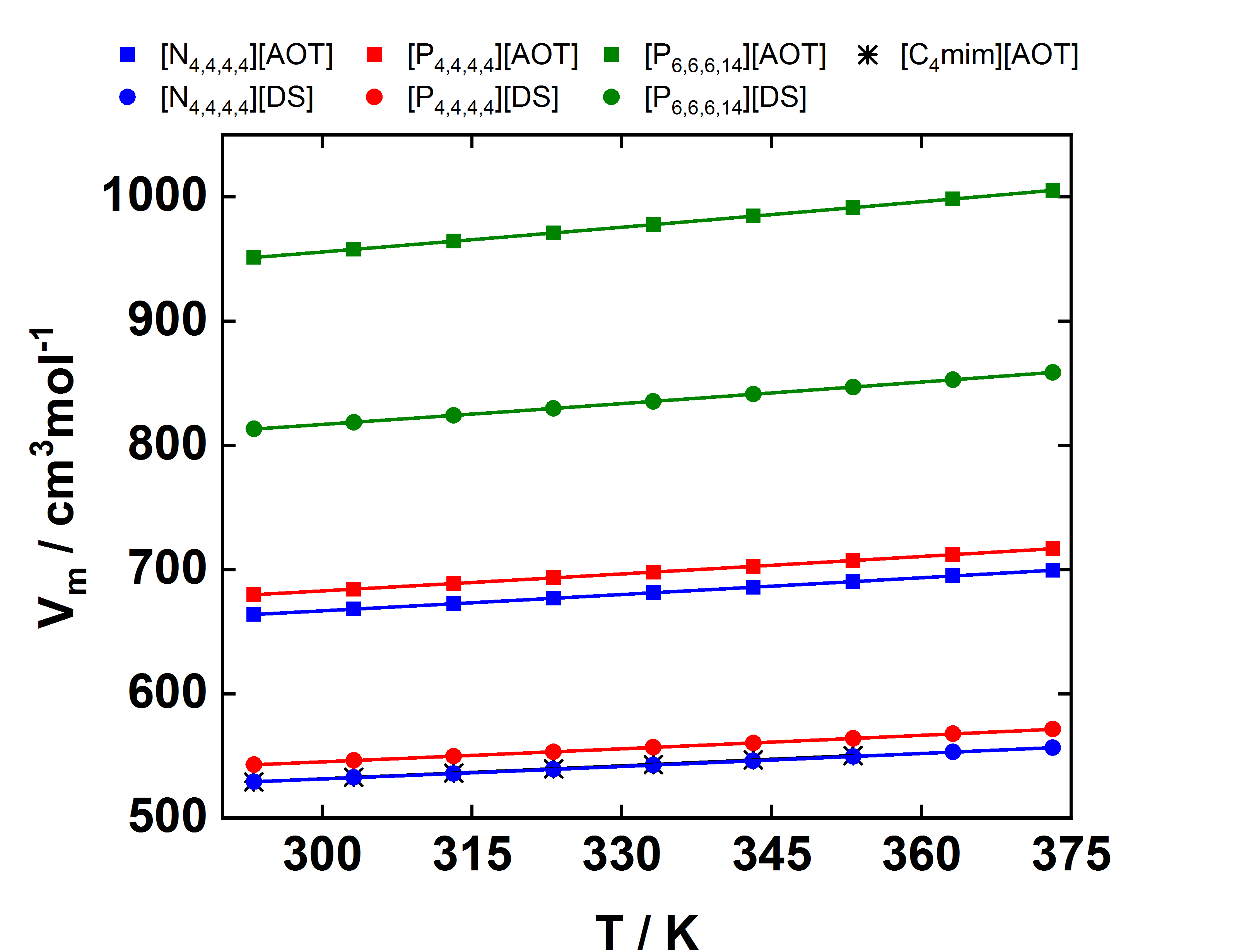}
    \caption{Molar volume of the ionic liquids depicted in Fig.\ref{fig:Struc_SAILs} as a function of temperature at ambient pressure.}
    \label{fig:Vm}
\end{figure}

The densities of the synthesized ionic liquids, measured in the temperature range \SIrange{293}{373}{\K}, are listed in Table S2 and expressed in molar volumes in Figure \ref{fig:Vm}. As expected, the molar volume slightly increases with increasing temperature for all the seven ILs due to thermal expansion. The effect of the temperature is similar for all the ionic liquids, as shown by the derivative of the fittings in Table S3.

Also as expected, the molar volume increases with increasing molecular weight of the ILs (Table S1), following the order \ce{[C4mim][AOT]} $\approx$ \ce{[N_{4,4,4,4}][DS]} < \ce{[P_{4,4,4,4}][DS]} < \ce{[N_{4,4,4,4}][AOT]} < \ce{[P_{4,4,4,4}][AOT]} < \ce{[P_{6,6,6,14}][DS]} < \ce{[P_{6,6,6,14}][AOT]}. Even though the imidazolium based ionic liquid is heavier than \ce{[N_{4,4,4,4}][DS]}, they present similar molar volumes because \ce{[C4mim][AOT]} is denser probably ought to a more compact liquid structure probably triggered by $\pi-\pi$ stacking.

As described in the literature, the molar volume of ionic liquids is additive and can be predicted by using a previously described group contribution method (GCM):\cite{Jacquemin2008,Deng2010a}
\begin{equation}
    \label{eq:GCM}
    \begin{split}
    V_{m}^{GCM} = \sum_j n_j\sum_{i=0}^2C_i(\delta T)^i
    \end{split}
\end{equation}
where $V_{m}^{GCM}$ denotes the calculated IL molar volume, $n_j$ is the number of groups listed in Table S4 for the seven ILs, $C_i$ are temperature dependent coefficients listed in Table S5, and $\delta T = T - 298.15 K$ with $C_0$ corresponding to the group's molar volume at \SI{298.15}{\K} and \SI{0.1}{\MPa}.

The experimental molar volume of the ammonium and phosphonium based ionic liquids paired with both \ce{[AOT]-} and \ce{[DS]-} anions can be predicted using the GCM within 2\% for temperatures lower than \SI{333}{\K} and within 3\% from \SI{333}{\K} up to \SI{373}{\K}, as listed in Table S2. For \ce{[C4mim][AOT]}, the calculated molar volume agrees to within 4\% with the experimental values thus confirming that the known additivity of the molar volume of ionic liquids is also observed for SAILs. The experimental molar volume is, nevertheless, systematically larger than the  $V_{m}^{GCM}$ for ILs containing the  \ce{[AOT]-} anion (\ce{[C4mim][AOT]}, \ce{[N_{4,4,4,4}][AOT]} and \ce{[P_{6,6,6,14}][AOT]}) and lower than $V_{m}^{GCM}$ for ILs containing \ce{[DS]-} anion (\ce{[N_{4,4,4,4}][DS]}, \ce{[P_{4,4,4,4}][DS]} and \ce{[P_{6,6,6,14}][DS]}). In view of the observed additivity of the molar volume for SAILs, these differences suggest that the linear and less voluminous anion, \ce{[DS]-}, packs into a more compact structure when compared to the branched and more voluminous \ce{[AOT]-} anion. 

\begin{figure}[hbt!]
    \centering
    \includegraphics[width=0.5\textwidth]{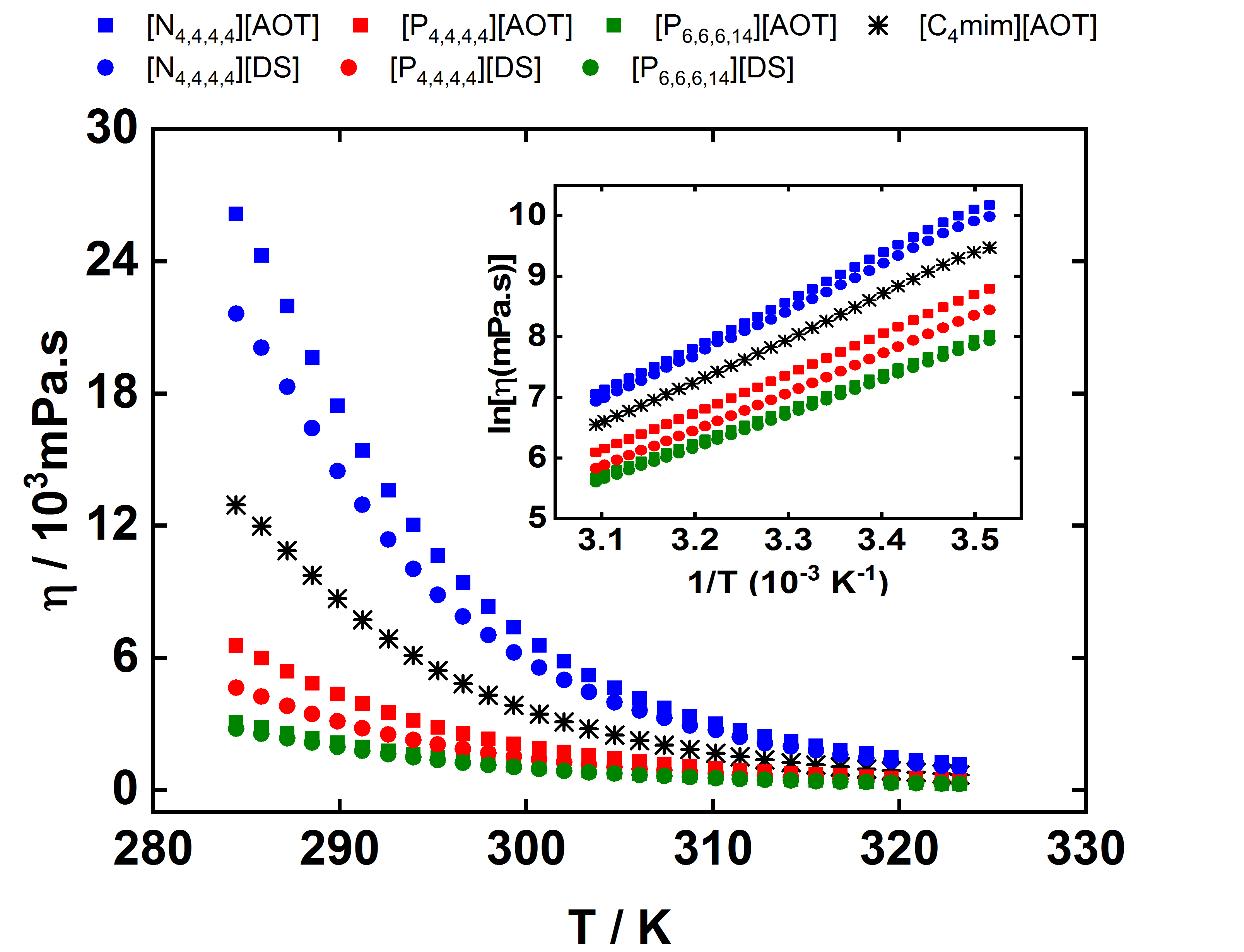}
    \caption{Viscosity as a function of temperature at a fixed shear rate of \SI{50}{\per\second}. The inset shows the fit to the Arrhenius equation \ref{eq:Arr_visc}.}
    \label{fig:Visc}
\end{figure}

The viscosity of the ionic liquids were measured in the temperature range \SIrange{283}{323}{\K}, the experimental data being listed in Table S6 and depicted in Fig.\ref{fig:Visc}. 
The viscosities of the SAILs reported herein vary between $2000$ and \SI{15000}{\m\Pa\s} at room temperature, which are generally two to four orders of magnitude higher than conventional ILs and organic solvents.\cite{Khalil2020, Alcalde2015, Yu2012} Conventional ILs are viscous due to electrostatic,  hydrogen bonding, van der Waals forces, and solvophobic interactions.\cite{Rooney2010, Okoturo2004} The higher viscosity of SAILs is attributed to interdigitation of their much longer alkyl chains.\cite{Zhang2022, Mao2019} For SAILs with the same anion the viscosity decreases in the order: \ce{[N_{4,4,4,4}]+} > \ce{[P_{4,4,4,4}]+} > \ce{[C4mim]+} > \ce{[P_{6,6,6,14}]+}. This order is mainly attributed to cations with localised charged centres and short alkyl chains interacting more strongly electrostatically with anions.\cite{Okoturo2004} For the same cations, SAILs with \ce{[AOT]-} anions are more viscous than those with \ce{[DS]-} anions, likely because \ce{[AOT]-} is larger, and has branched alkyl chains, which leads to steric hindrance. 



For all SAILs, the viscosity decreases sharply with temperature (Fig. \ref{fig:Visc}), which is in accordance with the behaviour observed for conventional ILs. Both Arrhenius and its modified version, Vogel--Fulcher--Tamman (VFT) equations, c.f. equation \ref{eq:Arr_visc} and equation (S8) are widely used to fit the viscosity data as a function of temperature. Okoturo et al.\cite{Okoturo2004} found that ILs with asymmetric cations tend to follow the Arrhenius law1, whereas ILs with symmetrical cations tend to follow the VFT equation.

For the SAILs investigated in this study, the variation of viscosity with temperature is well fitted by the following Arrhenius equation ($R^{2} > 0.999$), as shown in the inset of Fig. \ref{fig:Visc}:

\begin{equation}
    \label{eq:Arr_visc}
    \begin{split}
        \ln\eta\ = \ln\eta_{\infty}+ \frac{E_\eta}{RT}
    \end{split}
\end{equation}
where $E_{\eta}$ is the activation energy for the viscous flow, $\eta^{\infty}$ is the viscosity at the infinite temperature, and $R$ is the universal gas constant.\cite{Khan2017} The fitted parameters are listed in Table S7. $E_{\eta}$ describes the energy barrier for ions to move past each other, and thus indicates overall interactions between cations and anions. The fitted $E_{\eta}$ values for SAILs, which range from \SI{46}{\kilo\J\per\mol} to \SI{63}{\kilo\J\per\mol}, increase in the same order with the viscosity and are higher than most conventional ILs.\cite{Okoturo2004, Liu2011a, Jacquemin2006} These results are consistent with more energy being required for to move the large SAIL ions to past one another due to larger ionic dimensions, stronger van der Waals interactions, and alkyl chain interdigitation. 

The VFT equation is also used to fit the viscosity vs temperature data and is shown in  Fig. S1 and Table S7. 


\begin{figure}[hbt!]
    \centering
    \includegraphics[width=0.5\textwidth]{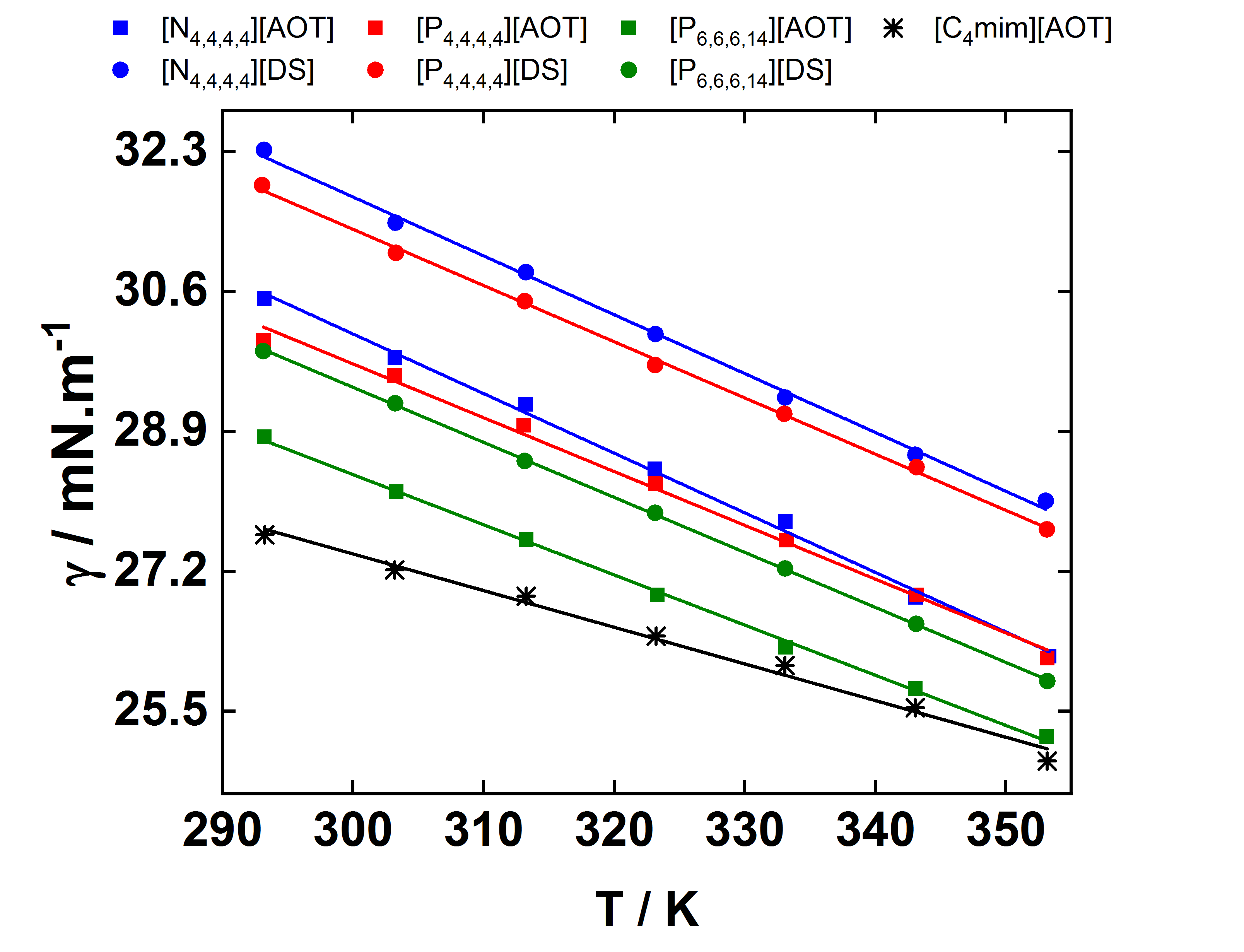}
    \caption{Surface tension of the ionic liquids depicted in Fig.\ref{fig:Struc_SAILs} as a function of temperature at air and ambient pressure.}
    \label{fig:STF}
\end{figure}

Figure \ref{fig:STF} and Table S8 shows the experimental data-sets of the surface tension $\gamma$ for the seven SAILs of this work. The surface tension of the pure ILs increases in the order \ce{[C4mim][AOT]} < \ce{[P_{6,6,6,14}][AOT]} < \ce{[P_{6,6,6,14}][DS]} < \ce{[P_{4,4,4,4}][AOT]} < \ce{[N_{4,4,4,4}][AOT]} < \ce{[P_{4,4,4,4}][DS]} < \ce{[N_{4,4,4,4}][DS]}, meaning that stronger cohesive forces are present in the liquid phase of the ILs based on the \ce{[DS]-} anion. This is especially observed when \ce{[DS]-} is paired with the short and symmetric alkyl phosphonium or ammonium cations, probably due to the most efficient packing of the ions. As expected, the surface tension decreases fairly linearly with increasing temperature. The values have been fitted to the empirical equation proposed by MacLeod \cite{Macleod1923}, Equation S1. The product of the constant of proportionality from Equation S1, $K$, by the molar mass, $M$, is known as the Parachor the compound $P_{ch} = KM$ \cite{Sugden1924}. 
Parachors obtained from the experimental $\gamma$ measurements of this work are presented in Table \ref{tbl:Pch}, together with comparisons from the predicted values using the method proposed by \citeauthor{Ericksen2002} \citep{Ericksen2002}, the absence of any other surface tension data hampering any other comparisons. This model is based on the fact that the Parachor of a substance can be estimated through a group contribution method. From the data available for about 400 compounds at the DIPPR database with surface tension measurement uncertainty less than 1 \%, \citeauthor{Ericksen2002}\citep{Ericksen2002} developed the group contributions values used here for the calculated Parachors of Table 1. As can be seen, our experimental values are always below the predicted ones, with similar relative deviations, yielding an average absolute deviation $AAD = 6.6 \%$. This result is in close agreement with the $AAD = 5.8 \%$ observed by \citeauthor{Gardas2008}\citep{Gardas2008} that studied 38 imidazolium-based ionic liquids with several different anions, following the same procedure. 

\begin{table}
\begin{center}
\small
  \caption{Parachors, $P_{ch, exp}$, obtained from the experimental surface tension data sets for the ionic liquid-based surfactant depicted in Fig.\ref{fig:Struc_SAILs} and comparison with the calculated Parachors, $P_{ch, calc}$, from Eq. S1. $\delta$ = $10^2 (P_{ch, exp} - P_{ch, calc})/P_{ch, calc}$.}
  \label{tbl:Pch}
    \begin{tabular}{lcccc}
    \toprule
                      & $M$ / $\si{\g\per\mol}$ & $P_{ch, exp}$ & $P_{ch, calc}$ & $\delta$ \\
    \midrule
    {\ce{[C4mim][AOT]}}   & 560.82     & 1225     & 1337      & -8               \\
    {\ce{[N_{4,4,4,4}][AOT]}}    & 664.06     & 1562     & 1679      & -7       \\
    {\ce{[P_{4,4,4,4}][AOT]}}    & 681.03     & 1598     & 1720      & -7        \\
    {\ce{[P_{6,6,6,14}][AOT]}}   & 905.50     & 2212     & 2359      & -6        \\
    {\ce{[N_{4,4,4,4}][DS]}}     & 507.85     & 1263     & 1337      & -6        \\
    {\ce{[P_{4,4,4,4}][DS]}}     & 524.82     & 1292     & 1378      & -6        \\
     {\ce{[P_{6,6,6,14}][DS]}}    & 749.29     & 1906     & 2017      & -5       \\
    \bottomrule
    \end{tabular}
\end{center}
\end{table}

The surface tension data also allowed to estimate the critical temperature, $T_c$, at the end point of a pressure \textit{vs} temperature phase equilibrium curve where the phase boundaries vanish, of all SAILs (Table \ref{tbl:Tc_Tb}). $T_c$ is commonly obtained by measuring the vapor pressure of a substance while increasing the temperature. This is particularly difficult for ionic liquids because: $(i)$ ionic liquids have a rather low vapor pressure making precise measurements difficult; and $(ii)$ most ionic liquids decompose before reaching the critical point. An approach to overcome these difficulties and estimate the critical temperature, $T_c$, of ionic liquids was proposed by \citeauthor{Rebelo2005}\citep{Rebelo2005} based on the temperature dependence of the density and surface tension. One of the best known empirical correlations of $\gamma$ versus $T$ is the Eötvos rule \cite{Shereshefsky1931}, Equation S2, and Guggenheim rule \cite{Guggenheim1945}, Equation S3, which take into account that $\gamma$ vanishes at the critical point and assumes that $\gamma$ behaves linear with $T$. 


\citeauthor{Rebelo2005}\citep{Rebelo2005} demonstrated the adequacy of this method by the assessment of experimental and estimated critical, $T_c$, and normal boiling, $T_b$, temperatures of an extensive data-sets for 90 compounds, including molten inorganic salts, hydrogen-bonded and non hydrogen-bonded organics and inorganics. They found that both $T_c$ and $T_b$ can be predicted with an error better than 10 \%, apart from strongly hydrogen-bonded substances. 
The estimated $T_c$ from the fitting to Equations S2 and S3 of experimental density and surface tension measurements for the SAILs of this work are reported in Table \ref{tbl:Tc_Tb}, together with $T_b$ deduced applying $T_b = 0.6T_c$, which is valid for most substances. The relative differences between the predictions from both Equations are in close agreement with deviations of less than 1.7 \%. This can be seen as a test of the internal consistency of the method, and the good accuracy of the derived values for the critical temperature. 
The exception is for the imidazolium sample, \ce{[C4mim][AOT]}, where a difference of nearly 7 \% is found between the Eötvos and the Guggehheim estimates. However, this was reported before by \citeauthor{Rebelo2005}\citep{Rebelo2005} when testing the method on imidazolium-based ionic liquids: decreasing the length of the alkyl chain of cation not only increases the predicted critical temperatures, but also results in larger discrepancies between the Eötvos and the Guggenheim equations. 

\begin{table}
\begin{center}
\small
  \caption{Estimated critical temperatures, $T_c$, and normal boiling temperatures, $T_b$ = 0.6 $T_c$, for the ionic liquid-based surfactant depicted in Fig.\ref{fig:Struc_SAILs}. $\delta$ = {$10^2 (T_{Eotvos} - T_{Guggenheim})/T_{Guggenheim}$}.}
  \label{tbl:Tc_Tb}
    \begin{tabular}{lccccc}
    \toprule
    \addlinespace[0.5ex]
    & \multicolumn{2}{c}{$T_c$ / $\si\K$} & \multicolumn{2}{c}{$T_b$ / $\si\K$} & $\delta$ \\
                                 & Eöt$^a$ & Gug$^b$    & Eöt$^a$     & Gug$^b$  & \\
    \midrule
    {\ce{[C_{4}mim][AOT]}}       & 1119       & 1047              & 672            & 628             & 6.9 \\
    {\ce{[N_{4,4,4,4}][AOT]}}    & 796        & 802               & 477            & 481             & -0.8 \\
    {\ce{[P_{4,4,4,4}][AOT]}}    & 856        & 850               & 513            & 510             & 0.6  \\
    {\ce{[P_{6,6,6,14}][AOT]}}   & 878        & 864               & 527            & 518             & 1.7  \\
    {\ce{[N_{4,4,4,4}][DS]}}     & 834        & 838               & 500            & 503             & -0.4 \\
    {\ce{[P_{4,4,4,4}][DS]}}     & 858        & 856               & 515            & 514             & 0.3 \\
    {\ce{[P_{6,6,6,14}][DS]}}    & 837        & 833               & 502            & 500             & 0.5 \\
    \bottomrule
    \end{tabular}
\end{center}
$^a$ Eq. S2,
$^b$ Eq. S3. Eöt and Gug stand for Eötvos and Guggenheim, repectively.
\end{table}
\renewcommand{\arraystretch}{1}

\begin{figure}[hbt!]
    \centering
    \includegraphics[width=0.5\textwidth]{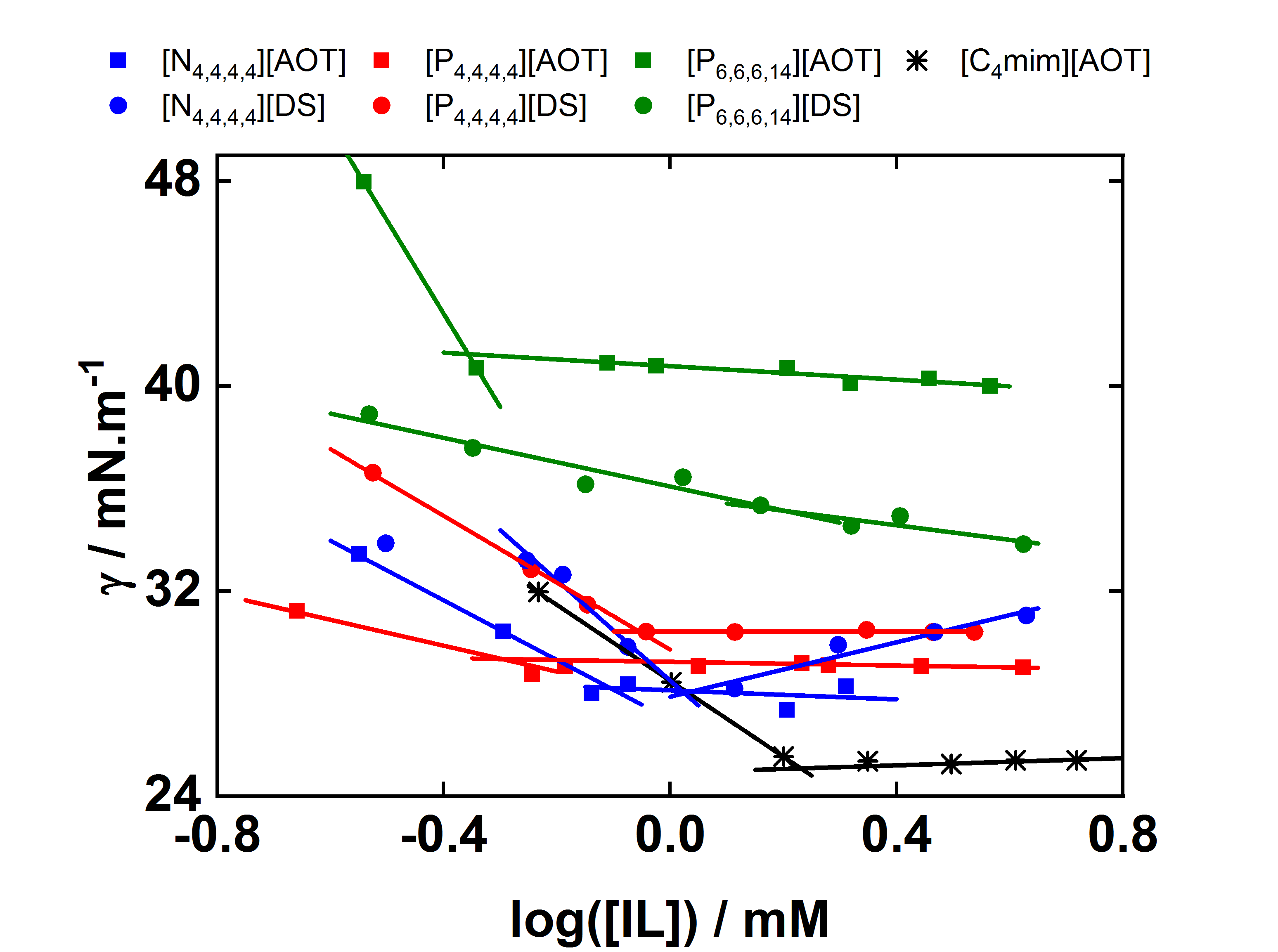}
    \caption{Experimental surface tension, $\gamma$, in aqueous solutions as a function of the logarithm of the surfactant ionic liquids of Figure \ref{fig:Struc_SAILs} concentration at the temperature $T$ = \SI{298.15}{\K} in air at ambient pressure.}
    \label{fig:cmc_SAILs}
\end{figure}

To evaluate the surfactant character of the SAILs, surface tension measurements were carried out also in ILs/water mixtures. Table S9 and Fig. \ref{fig:cmc_SAILs} show the sample composition and surface tension data of the mixtures. All the ionic liquids are effective in reducing the surface tension of the solvent and a critical micelle concentration (cmc) was found at concentrations lower than 2 mM for all the mixtures (Table \ref{tbl:cmc}). Except for \ce{[C4mim][AOT]}, the cmc is lower for the ILs bearing \ce{[AOT]-} anion and decrease in the order \ce{[P_{6,6,6,14}]+} > \ce{[N_{4,4,4,4}]+} > \ce{[P_{4,4,4,4}]+} in both \ce{[AOT]-} and \ce{[DS]-} based salts.

\begin{table}[htb]
\small
\begin{center}
    \caption{Critical micelle concentration, $cmc$, surface tension at the cmc, $\gamma_{cmc}$, and the effectiveness of the surface tension reduction, $\gamma_0-\gamma_{cmc}$ of the SAILs (Figure \ref{fig:Struc_SAILs}) in dilute aqueous solutions at $T$ = \SI{298.15}{\K} from surface tension measurements in air at ambient pressure. $\delta$ = $10^2 (cmc_{exp} - cmc_{lit})/cmc_{lit}$.}
    \label{tbl:cmc}
        \begin{tabular}{lcccc}
        \toprule
        \addlinespace[0.5ex]
        IL & $\displaystyle\frac{cmc}{mM}$ & $\displaystyle\frac{\gamma_{cmc}}{ \si{\mN\per\m}}$ & $\delta$ & $\displaystyle\frac{\gamma_0-\gamma_{cmc}}{\si{\mN\per\m}}$\\
        \midrule
        \multicolumn{5}{c}{dilute aqueous solutions} \\
        \ce{[C4mim][AOT]}           & 1.70 & 25.08 & $-4.4^{a,1}$ & 42.23 \\  
        \ce{[N_{4,4,4,4}][AOT]}     & 0.78 & 28.22 & $-2.1^{b,2}$ & 45.37 \\  
                                    &      &       & $1.7^{b,1}$  &  \\ 
                                    &      &       & $-0.6^{c,2}$ &  \\ 
                                    &      &       & $0.5^{c,1}$ &   \\ 
        \ce{[P_{4,4,4,4}][AOT]}     & 0.51 & 29.34 &  & 41.11\\
        \ce{[P_{6,6,6,14}][AOT]}    & 0.88 & 40.53 &  & 29.92\\
        \ce{[N_{4,4,4,4}][DS]}      & 1.06 & 28.01 &  & 42.44\\
        \ce{[P_{4,4,4,4}][DS]}      & 0.88 & 30.42 &  & 40.03\\
        \ce{[P_{6,6,6,14}][DS]}     & 1.60 & 35.12 &  & 35.33\\
        \bottomrule
        \end{tabular}
\end{center}
$^a$ Brown et al. 2012 \cite{Brown2012} \\
$^b$ Chakraborty et al. 2008 \cite{Chakraborty2008}\\
$^c$ Kumar et al. 2021 \cite{Kumar2021} \\
$^1$ From surface tension measurements \\
$^2$ From electrical conductivity measurements
\end{table}

\begin{figure}[hbt!]
    \centering
    \includegraphics[width=0.5\textwidth]{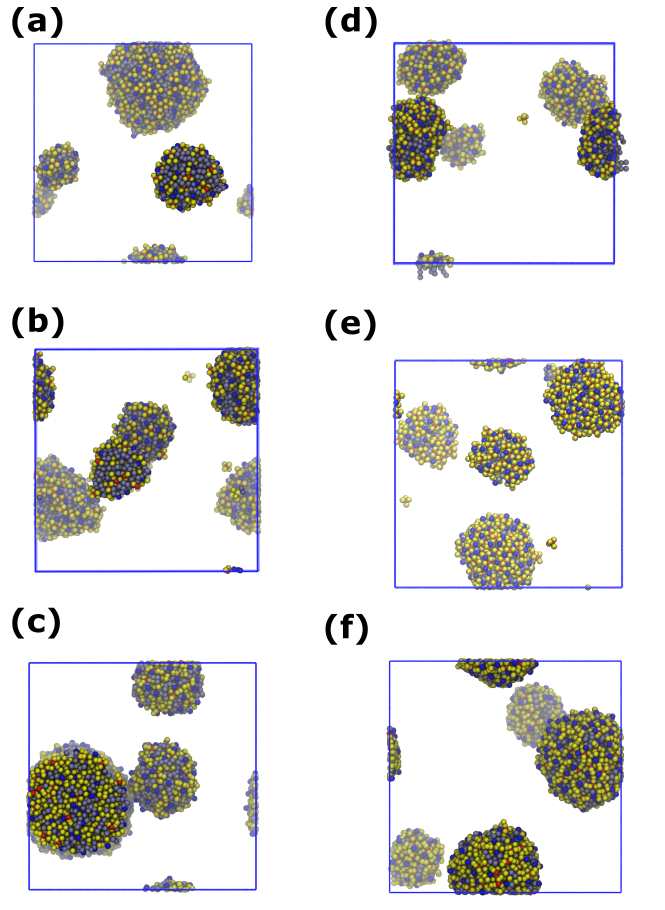}
    \caption{ Snapshot of the simulations of 0.1 M SAIL aqueous solutions at \SI{298.15}{\K}. Where the cations are represented in red (headgroup) and yellow, while the anions in dark (headgroup) and light blue. (a) \ce{[N_{4,4,4,4}][AOT]}, (b) \ce{[P_{4,4,4,4}][AOT]}, (c) \ce{[P_{6,6,6,14}][AOT]}, (d) \ce{[N_{4,4,4,4}][DS]}, (e) \ce{[P_{4,4,4,4}][DS]}, and (f) \ce{[P_{6,6,6,14}][DS]}. }
    \label{fig:sim_micelles}
\end{figure}

Coarse-grain molecular dynamics simulations allowed for the further characterization of the micelle formation of the SAILs in aqueous solutions. For solutions of 0.1 M at 298.15 K, spherical micelles were observed in all systems, as illustrated in Figure \ref{fig:sim_micelles}. In Table \ref{tbl:micelle_size}, the maximum micelle size in terms of number of ion pairs is presented. The trends for the micelle sizes follow the same observed for the critical micelle concentrations in terms of the anions, with the lowest cmc's corresponding to the largest micelles. 

\begin{table}[hbt!]
\begin{center}
\small
 \caption{\ The maximum observed micelle size defined in terms of the number of ion pairs within the micelle, for SAILs 0.1 M aqueous solutions at \SI{298.15}{\K}.}
 \label{tbl:micelle_size}
 \begin{tabular}{lc}
 \toprule
 IL &  Maximum micelle size \\
 \midrule
 \ce{[N_{4,4,4,4}][AOT]}  & 314 \\
 \ce{[P_{4,4,4,4}][AOT]}  & 260 \\
 \ce{[P_{6,6,6,14}][AOT]} & 308 \\
 \ce{[N_{4,4,4,4}][DS]}   & 219 \\
 \ce{[P_{4,4,4,4}][DS]}   & 173 \\
 \ce{[P_{6,6,6,14}][DS]}  & 183 \\
 \bottomrule
  \end{tabular}
 \end{center}
\end{table}

The micellization parameters of the SAILS were then derived from the surface tension measurements at different IL concentrations in the SAILs/water mixtures.\cite{Martinez-Landeira2002} The minimum molecular area at the air-liquid interface, $A_{mim}$, gives insights on the packing density of the solute at the interface and it was calculated by assuming ideality from the Gibbs equation, Equation S4. 
The standard free energy of micellization, $\Delta{G}_{mic}^0$, was estimated by Equation S6. 

\begin{figure}[hbt!]
    \centering
        \includegraphics[width=0.55\textwidth]{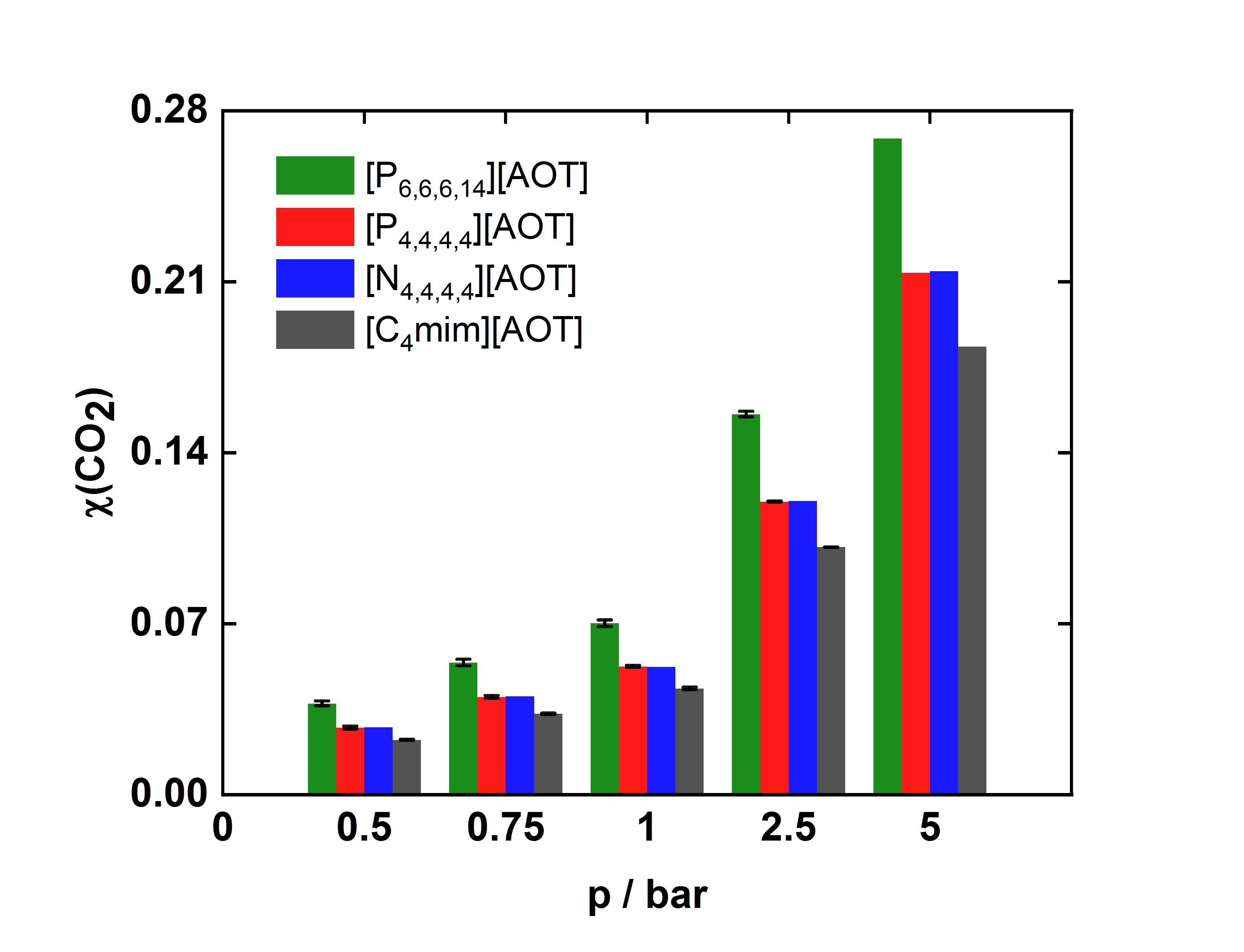}
        \includegraphics[width=0.55\textwidth]{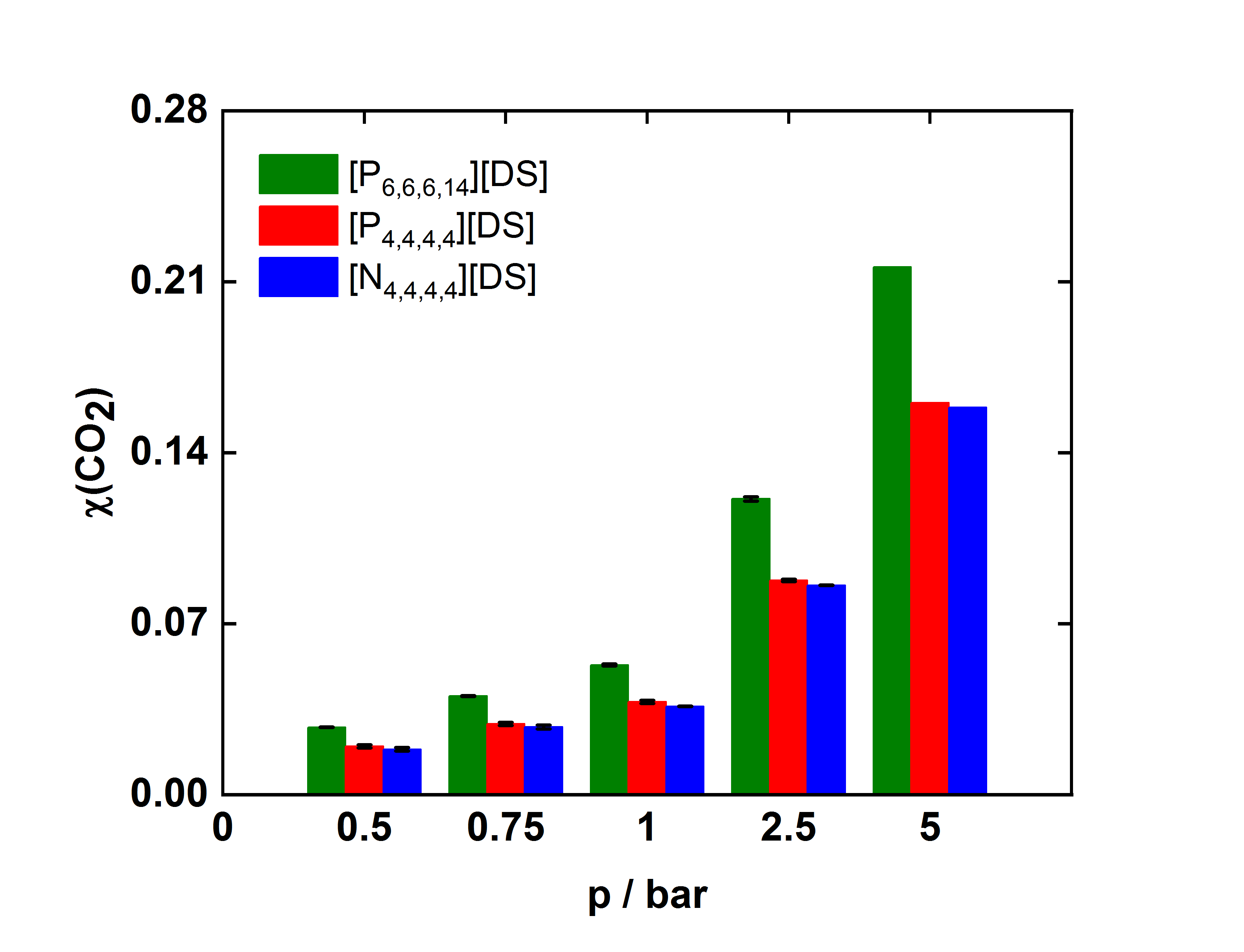}
    \caption{\ce{CO2} average absorption by the SAILs depicted in Fig.\ref{fig:Struc_SAILs} as a function of pressure at \SI{303}{\K}. The error bars correspond to the deviations in the absorption/desorption cycle.}
    \label{fig:DS_gasCO2}
\end{figure}

\begin{figure}[hbt!]
    \centering
        \includegraphics[width=0.55\textwidth]{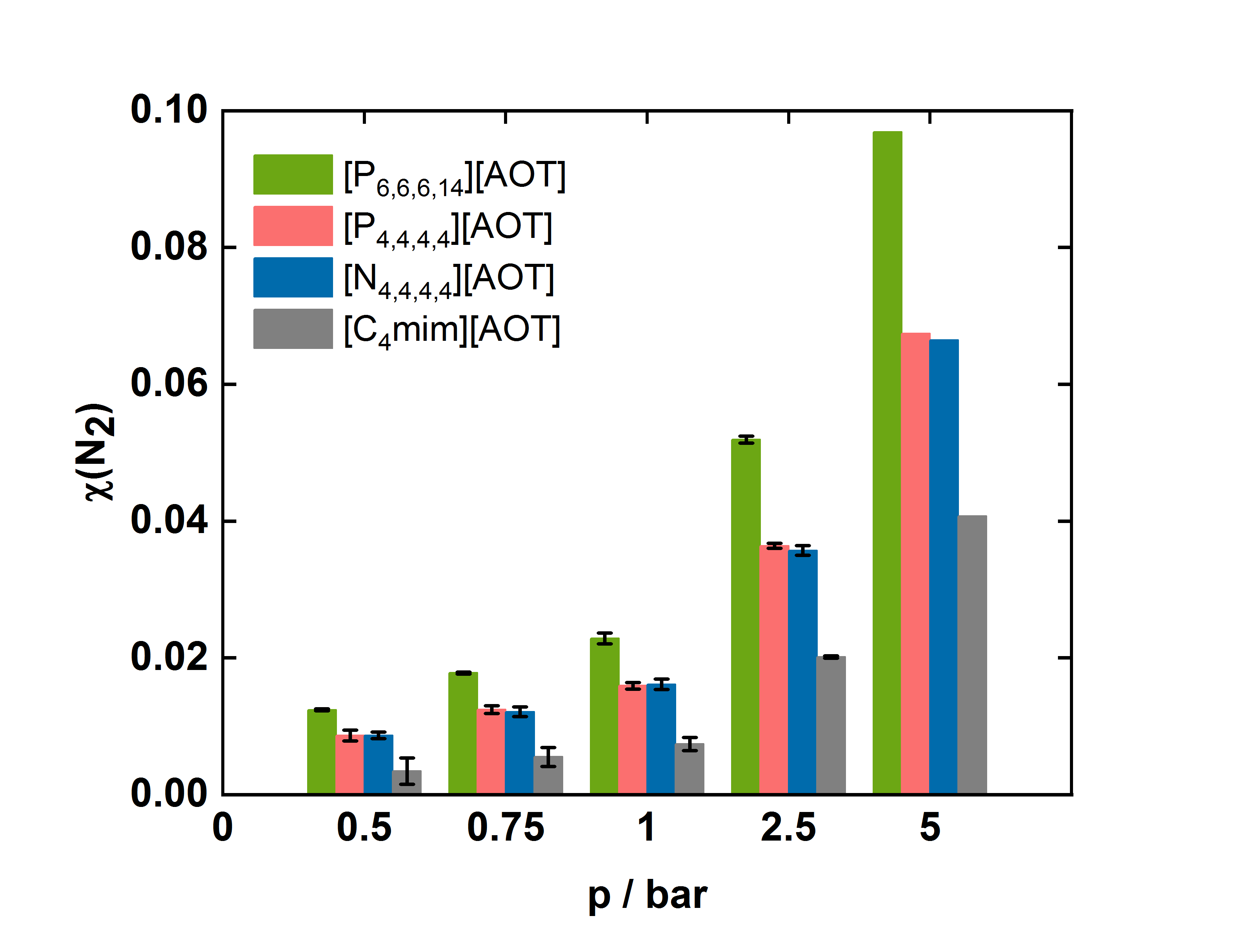}
        \includegraphics[width=0.55\textwidth]{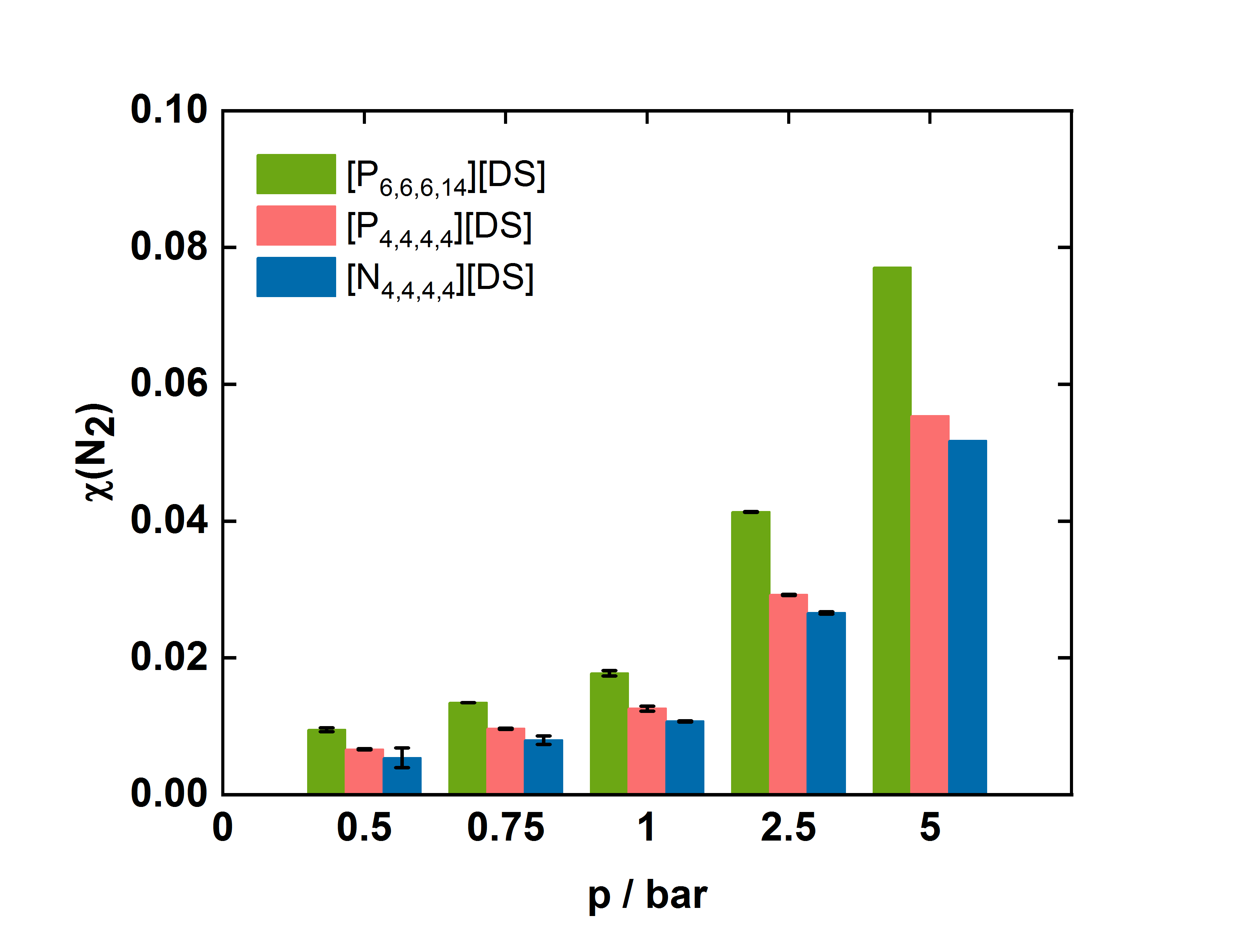}
    \caption{\ce{N2} average absorption by the SAILs depicted in Fig.\ref{fig:Struc_SAILs} as a function of pressure at \SI{303}{\K}. The error bars correspond to the deviations in the absorption/desorption cycle. }
    \label{fig:DS_gasN2}
\end{figure}

The standard free energy of adsorption, $\Delta{G}_{ad}^0$, describes the free energy change for the adsorption of amphiphiles to the liquid-air interface (see Equation S7). The surface tension of the neat solvent, $\gamma_0$, is $70.45\pm{0.28}$ \si{\mN\per\m} (Table \ref{tbl:cmc}). The surface tension of water was measured at the same conditions as the SAILs aqueous solutions and the fluctuations of the surface tension with the repetition are depicted in Fig. S2. 

The results depicted in Table \ref{tbl:micelle_param} show that the free energy of association is negative for all the SAILs, indicating the micelle formation is a spontaneous processes. Except for \ce{[C4mim][AOT]}, $\Delta{G}_{mic}^0$ is more negative for the ionic liquids bearing \ce{[AOT]-} anion, meaning that micelles are more easily formed and more stable than in the ionic liquids based on the \ce{[DS]-} anion. The calculated free energy of adsorption is more negative than the free energy of micellization for all SAILs, which means the adsorption of the ILs at the interface is more spontaneous than micelle formation and that the micelles are formed after the adsorption at the interface. 

The maximum surface excess, also depicted in Table \ref{tbl:micelle_param} is positive for all the SAILs, indicating the concentration of the IL is higher at the liquid-air interface than in the bulk because the hydrophobic chains prefers to be at liquid-air interface such as a typical surface active monomer.\cite{Kumar2021a} Except for \ce{[C4mim][AOT]}, $A_{min}$ is smaller for the ionic liquids with the \ce{[DS]-} anion, meaning they present higher packing density, thus highest hydrophobic interaction in water than the ionic liquids based on the \ce{[AOT]-} anion. The $A_{min}$ obtained for \ce{[P_{6,6,6,14}][AOT]} is highlighted in italic on Table \ref{tbl:micelle_param} as it is a less reliable value as due to the lower cmc, only two points could be measured below this concentration. The highest $A_{min}$ of \ce{[P_{4,4,4,4}][AOT]} and \ce{[P_{4,4,4,4}][DS]} when compared to \ce{[N_{4,4,4,4}][AOT]} and \ce{[N_{4,4,4,4}][DS]} indicate weaker hydrophobic interactions in the phosphonium based ionic liquids. Brown \textit{et al.}\cite{Brown2011} reported a value of $A_{min}$ for \ce{[N_{3,3,3,3}][DS]} smaller than \ce{[N_{3,3,3,3}][AOT]} and explained this as an effect of the higher tail volume of the \ce{[AOT]-} anion on the calculation of the packing parameter.

\begin{table}[hbt!]
\begin{center}
\small
 \caption{\ The maximum surface excess, $\Gamma_{max}$, and the minimum molecular area, $A_{mim}$, of the SAILs at the interface of SAILs/water mixtures; the standard Gibbs free energy of micellization, $\Delta{G}_{mic}^0$, and of adsorption, $\Delta{G}_{ad}^0$, at \SI{298}{\K}.}
 \label{tbl:micelle_param}
 \begin{tabular}{lcccc}
 \toprule
 IL &  $\displaystyle\frac{\Gamma_{max}}{10^{-6}\si{\mol\m}^{-2}}$ & $\displaystyle\frac{A_{mim}}{\si{\square\angstrom\per\molec}}$ & $\displaystyle\frac{\Delta{G}_{mic}^0}{\si{\kJ\per\mol}}$ &  $\displaystyle\frac{\Delta{G}_{ad}^0}{\si{\kJ\per\mol}}$ \\
 \midrule
 \ce{[C4mim][AOT]}        & 2.6 & 64.1 & -15.8   & -33.9 \\
 \ce{[N_{4,4,4,4}][AOT]}  & 2.0 & 81.5 & -17.7  & -39.2 \\
 \ce{[P_{4,4,4,4}][AOT]}  & 1.0 & 160.1 & -18.8  & -59.9 \\
 \ce{[P_{6,6,6,14}][AOT]} & 3.0 & \textit{55.5} & -17.5   & -28.0 \\
 \ce{[N_{4,4,4,4}][DS]}   & 3.4 & 48.5 & -17.0   & -29.9 \\
 \ce{[P_{4,4,4,4}][DS]}   & 2.3 & 72.6 & -17.4   & -35.6 \\
 \ce{[P_{6,6,6,14}][DS]}  & 0.8 & 198.7 & -16.0  & -60.1 \\
 \bottomrule
  \end{tabular}
 \end{center}
\end{table}

Figures \ref{fig:DS_gasCO2}-\ref{fig:DS_gasN2} and Tables S10-S13 show the \ce{CO2} or \ce{N2} solubility in all synthesized SAILs. The gas absorption of either \ce{CO2} or \ce{N2} follow the molar volume of the ionic liquids and increase in the order \ce{[N_{4,4,4,4}][DS]} = \ce{[P_{4,4,4,4}][DS]} < \ce{[N_{4,4,4,4}][AOT]} = \ce{[P_{4,4,4,4}][AOT]} <  \ce{[P_{6,6,6,14}][DS]} < \ce{[P_{6,6,6,14}][AOT]}. In the \ce{[AOT]-}-based ionic liquids the gas absorption also follows the molar volume, increasing in the order \ce{[C4mimm][AOT]} < \ce{[N_{4,4,4,4}][AOT]} = \ce{[P_{4,4,4,4}][AOT]} < \ce{[P_{6,6,6,14}][AOT]}. Both anion and cation play a part in the gas absorption, the highest solubility being that of \ce{CO2} in \ce{[P_{6,6,6,14}][AOT]}, about 0.3 in mole fraction. The branched \ce{[AOT]-} anion and the asymetric and longer alkyl chains present in the \ce{[P_{6,6,6,14}]+} cation contributing to a larger gas absorption than the linear \ce{[DS]-} anion and the symmetric and shorter alkyl chains present in the \ce{[P_{4,4,4,4}]+} cation, respectively. The solubility of both \ce{CO2} and \ce{N2} is larger in the SAILs with larger non-polar domains, all phosphonium and ammonium based ILs absorbing more the gases than the imidazolium based one, which can be explained by the more favourable entropy of solvation. Changing the phosphorous by a nitrogen atom does not seem to affect the SAILs polar domain, as indicated by the similar \ce{CO2} and \ce{N2} solubilities. 

Taking the advantages of the ionic liquids to dissolve ionic salts, mixing with both polar and non-polar molecular solvents and  their capacity of absorbing gases, Zhang \textit{et. al}\cite{Zhang2015} proposed low viscosity IL/surfactant aqueous solutions to be used in \ce{CO2} capture. Their results showed an improvement in the \ce{CO2} absorption capacity when compared with the pure surfactant aqueous solutions. Therefore, the gas absorption in these new SAILs aqueous solution are subject of further studies.

\section{Conclusion}

The effect of ionic structure on physicochemical properties and gas adsorption has been investigated systematically for a series of SAILs with imidazolium, phosphonium, and ammonium cations paired with \ce{[AOT]-} or \ce{[DS]-} anions. SAILs including linear \ce{[DS]-} anions pack more compactly than SAILs with branched \ce{[AOT]-} anions, the additivity of the molar volumes still being verified by the calculation of the molar volumes of the salts using group contribution methods. Compared to conventional ILs, SAILs are more viscous and require higher energy for ions to move across each other owing to larger ionic dimensions and alkyl chain interdigitation. This results form a balance between the interdigitation of the alkyl side-chains of the ions, particularly important for \ce{[AOT]-} based SAILs, and the electostatic interactions between cations and anions, prevalent for the smaller cations \ce{[P_{4,4,4,4}]+} and \ce{[N_{4,4,4,4}]+}.

The surface tensions of SAILs are lower than those of conventional ILs and closer to those of n-alkanes, because SAIL ions orient their alkyl chains towards the air when adsorbed at the liquid-air interface following a surfactant-like behaviour. The surface tension dependence with the temperature allowed the estimation of the critical temperatures of all SAILs, overcoming the experimental difficult in obtaining these directly due to the low vapor pressure of ionic liquids. The critical temperatures are very similar for all SAILs at around \SI{900}{\K}. When dissolved in water, SAIL ions adsorb to the liquid-air interface spontaneously, and then form spherical micelles when the concentration of amphiphilic SAIL ions is higher than cmc, as per conventional surfactants in water. 

The \ce{CO2} and \ce{N2} absorption measurements reveal that both the anions and cations play a role in the increase of the gas solubilities with SAILs having longer cation alkyl tails and branched anions absorbing larger quantities of gas. This is observed for gases that do not interact specifically with the liquid solvent, their absorption being governed by van der Waals type interactions.

\section*{Conflicts of interest}
There are no conflicts to declare.

\section*{Acknowledgements}

 The authors thank IDEX-LYON for financial support (Programme Investissements d’Avenir ANR-16-IDEX-0005). D.L.M. thanks for the support to ERDF/“Junta de Castilla y León” (CLU-2019-04).



\balance




\bibliographystyle{rsc} 

\providecommand*{\mcitethebibliography}{\thebibliography}
\csname @ifundefined\endcsname{endmcitethebibliography}
{\let\endmcitethebibliography\endthebibliography}{}

\newpage

\end{document}